# Nucleate Boiling Simulation using Interface Tracking Method

**Mengnan Li[1], Joachim Moortgat[1], Igor A. Bolotnov*[2]**

[1] School of Earth Science, The Ohio State University, 125 Oval Dr S

Columbus, OH 43210, USA

[2] Department of Nuclear Engineering, North Carolina State University, 2500 Stinson Dr.,

Raleigh, NC 27695, USA

li.5382@osu.edu, Moortgat.1@osu.edu, iabolotn@ncsu.edu

**ABSTRACT**

The development and validation of 3D multiphase computational fluid dynamics (M-CFD) models and physics-informed data-driven modeling require data of high-quality and high-resolution. Considering the difficulties in acquiring the corresponding experimental data in prototypical conditions, two-phase boiling simulations by Interface Tracking Method (ITM) based models can be used to generate high-resolution numerical data in a consistent and relatively economical manner. A boiling model is developed in one of the ITM-based multiphase-flow solvers, named PHASTA, to investigate the nucleate boiling phenomenon. The interaction between bubbles forming at adjacent nucleation sites is investigated with this ITM boiling model. Nucleate pool boiling simulations with multiple nucleation sites are presented in this paper and influences of site distance, neighboring bubble size and contact angle effect are investigated. The presented boiling model can conduct boiling simulation on 3D unstructured computational meshes. These simulation results improve our understanding of the physical mechanisms of the nucleate boiling phenomenon and provide high-resolution numerical data for M-CFD validation and advanced boiling model development.

## 1. INTRODUCTION

Boiling, as one of the most efficient heat transfer mechanisms, is widely used in various engineering systems. Due to its complex nature, better understanding and modeling of the boiling process remains a major challenge in multiphase flow research. In Light Water Reactor (LWR) nuclear power plants, the distribution of vapor in the reactor core sub-channels affects the heat transfer rate and may cause unfavorable conditions, such as departure from nucleate boiling (DNB) phenomenon. The DNB later on may cause fuel cladding damage and lead to unplanned reactor shutdowns and accidents.

The interaction between bubbles forming at adjacent nucleation sites has a significant influence on the characteristics of the nucleate boiling process, like bubble release frequency, departure diameter and active nucleation site density [1-3], but in most nucleate boiling models this interaction is not considered [4,5] or relies on empirical correlations [6]. As the distance between neighboring nucleation sites changes, one nucleation site can either promote or inhibit the nucleate process of the nearby sites. Chekanov [7] was the first to experimentally study the interaction between nucleation sites with two artificial nucleate sites immersed in water. He proposed a dimensionless cavity spacing $S/D_d$ to categorize the interactions into three regions. $S$ is the distance between two cavity centers and $D_d$ is the average bubble departure diameter. For $S/D_d < 3$, the formation of a bubble at one nucleation site inhibits the formation of the neighboring bubble while for $S/D_d > 3$, the formation of a bubble promotes the formation of the bubble at the neighboring nucleate site. If $S/D_d \gg 3$, the bubble growth rates at neighboring nucleate sites are independent from each other. Calka and Judd [8] investigated the interaction phenomena at adjacent nucleation sites during saturated boiling of dichloromethane. Like Chekanov, they used a similar three-region approach to describe the interactions between nucleation sites. $S/D_d < 1$ is called the "promotive" region where the bubble at an adjacent nucleation site will form more frequently compared to those at single nucleation site. $1 < S/D_d < 3$ is an "inhibitive" region, where the bubble formation at one nucleation site has a negative effect on the bubble formation at an adjacent nucleation site. $S/D_d > 3$ is the "independent" region which is the same as in Chekanov's model. Bonjour et al. [9] suggested a map of nucleation site interactions, which allowed the determination of site activation and bubble coalescence conditions according to the parameters of an experiment with three nucleation sites. Mukherjee and Dhir [10] experimentally studied lateral merger of vapor bubbles. They found that the merger of multiple bubbles significantly increases the overall wall heat transfer. Theofanous et al. [11,12] conducted a series of boiling experiments under highly controlled conditions. Their experimental data provides quantitative information on nucleation site density and nucleate boiling heat transfer over a broad range of heat fluxes, from the onset of nucleate boiling to the occurrence of crisis.

Boiling experiments provide valuable insights into bubble interactions between adjacent nucleation sites, but high-resolution experimental measurements are hard, expensive and time-consuming to perform. With advanced computation resources, high-resolution simulations using interface tracking methods can provide detailed flow and interface dynamics information of nucleation site interactions to help develop nucleate boiling models with high accuracy. However, there are limited numerical simulations in the literature focusing on the interactions of neighboring bubbles. Mukherjee and Dhir [10] conducted 3D simulations of the lateral merger of vapor bubbles during nucleate pool boiling. Calculations were carried out for multiple bubble mergers in a line and in a plane using uniform structured meshes. Their results show that the bubble merger process increased the overall wall heat transfer by trapping a liquid layer between bubble bases and by drawing cooler liquid towards the wall during contraction. Sato and Niceno [13] performed a nucleate pool boiling simulation from the boiling regime of discrete bubbles to a vapor mushroom regime. Various hydrodynamic interactions between bubbles, such as lateral coalescence between adjacent bubbles, vertical coalescence between consecutive bubbles, etc., are observed in their simulations. Most boiling simulations in the literature are conducted in cuboid or tube-like domains with structured grids. Despite the reduced cost of computational resources, 3D boiling simulations with interface tracking, especially for engineering applications, remains relatively expensive. To support the development and validation of 3D multiphase computational fluid dynamics (M-CFD) model and physics-informed data-driven modeling, it is essential to increase the affordability of simulations by improving numerical methods and ITM boiling models.

The boiling simulations presented in this paper are conducted with local mesh refinement, unstructured grids, and a highly scalable algorithm for parallel computing. It can integrate the mechanism study of the local boiling phenomenon as a subgrid model and utilize this to study the quantities of interest in large scale simulation for engineering applications. This approach may help fill the numerical data gap between the study of local phenomena and large scale engineering problems by performing high-resolution high-quality large-scale boiling simulations in practical geometries.

## 2. NUMERICAL METHODS

### 2.1. PHASTA Overview

The three-dimensional finite-element based code for incompressible flows — PHASTA (<u>P</u>arallel, <u>H</u>ierarchic, higher-order accurate, <u>A</u>daptive, <u>S</u>tabilized, <u>T</u>ransient <u>A</u>nalysis) is used for the presented research. PHASTA is the first unstructured grid LES code [14]. Anisotropic adaptive algorithms [15] and LES/DES models [16] have also been utilized in PHASTA [17]. The FEM formulation ensures the support of unstructured meshes including tetrahedral and hexahedral shape volume mesh and tetrahedral, wedge, or hexahedral shape boundary layer mesh[18,19], which is an important advantage for simulations in complex engineering geometries, for example, a PWR fuel subchannel with spacer grids and mixing vanes [20]. Moreover, it has high scalability on supercomputers (e.g., up to 3 million partitions using a 92 billion element mesh [21]). It has been demonstrated that PHASTA can generate high fidelity numerical data for the prediction of adiabatic single-phase flow phenomena [22]. After coupling with the level-set method, two-phase flow modeling becomes available in PHASTA [23]. A broad range of two-phase phenomena including the interaction between bubbles and flow [24], reactor bubbly flow [25] and two-phase heat transfer problems [26,27], bubble release frequency[28,29] can be simulated and investigated using PHASTA.

### 2.2. Flow Solver

The governing equations, interface tracking method, and boiling model are introduced in this section. In PHASTA, the transient incompressible Navier-Stokes (INS) equations are solved in three dimensions using a stabilized finite element method (FEM). The global system of INS equations is generated from a series of element-wise equations. Then a transformation of coordinates is made from the subdomains' local nodes to the domain's global nodes. This spatial transformation includes orientation adjustments in relation to the reference coordinate system. The temporal and spatial discretization of INS equations are provided by Whiting and Jansen [22]. The strong form of three conservation equations (continuity, momentum and energy) is shown below[16]:

$$\nabla \cdot \vec{u} = 0 \tag{1}$$

$$\rho \left[\frac{\partial \vec{u}}{\partial t} + (\vec{u} \cdot \nabla)\vec{u}\right] = -\nabla p + \mu \nabla^2 \vec{u} + \rho \vec{f} \tag{2}$$

$$\rho c_p \left(\frac{\partial T}{\partial t} + (\vec{u} \cdot \nabla)T\right) = \nabla \cdot (k\nabla T) + \vec{q} \tag{3}$$

where $\vec{u}$ denotes the fluid velocity, $\rho$ is the density of the fluid, $p$ is the static pressure, $\mu \nabla^2 \vec{u}$ is the viscous term and $\vec{f}$ is the body force density. $T$ is the absolute temperature of the fluid; $c_p$ is the specific heat at constant pressure ($c_p \approx c_v$ for incompressible flow) and $k$ the thermal conductivity. $\vec{q}$ denotes the dissipation function representing the work done against viscous forces, which is negligble in simulations of low-viscosity flow like water.

For incompressible flow, the component of viscous stress tensor($\mu \nabla^2 \vec{u}$) is expressed in terms of the liquid's viscosity and strain rate tensor of a Newtonian fluid as below:

$$\tau_{ij} = 2\mu S_{ij} = \mu(u_{i,j} + u_{j,i}) \tag{4}$$

The Continuum Surface Force (CSF) model of Brackbill et al. [30] is utilized to represent surface tension effect as a local interfacial force density.

$$\vec{f_s} = \sigma \kappa \delta(\phi) \vec{n} \tag{5}$$

where κ is the curvature of the interface, $\sigma$ is the surface tension.

### 2.3. Level-set Method

PHASTA uses the 'one-fluid' formulation to describe two-phase flow systems. The same set of conservation equations are utilized for the entire computational domian. Mulitiphase flow with sharp property variances across an interface are treated as a continous fluid whose properties vary from liquid phase to gas phase over a narrow range of values [31]. The interface between phases is resolved and tracked using numerical interface tracking methods.

As one of the widely-used interface tracking algorithms, the level-set method (LS) proposed by Sussman [32-35] and Sethian [36] is implemented in PHASTA. The LS method introduces a distance field (level-set field) $\varphi$ to store information about the distance to the interface. The interface is modeled as a zero level-set ($\varphi = 0$) of the smooth function. Distinct phases are denoted by the sign of the level-set field ($\varphi$): the liquid phase is represented by a positive value ($\varphi > 0$) while the gas phase is represented using a negative value ($\varphi < 0$). The advection of this level-set field $\varphi$ through the computational domain is given by the following equation:

$$\frac{D\varphi}{Dt} = \frac{\partial \varphi}{\partial t} + \vec{u} \cdot \nabla \varphi = 0 \tag{6}$$

where $\vec{u}$ is the local velocity vector.

The thickness of the mixed range is described using the number of grid elements across the interface ($\varepsilon$). The discontinuity at the interface is smoothed using a Heaviside kernel function, $H_\varepsilon$, in Eq. (7) to avoid numerical instabilities [33]:

$$H_\varepsilon(\varphi) = \begin{cases} 0 & , \varphi < -\varepsilon \\ \frac{1}{2}\left[1 + \frac{\varphi}{\varepsilon} + \frac{1}{\pi}\sin\left(\frac{\pi\varphi}{\varepsilon}\right)\right] & , |\varphi| < \varepsilon \\ 1 & , \varphi > \varepsilon \end{cases} \tag{7}$$

The formulation of flow density ($\rho(\varphi)$), viscosity ($\mu(\varphi)$), specific heat ($c_p(\varphi)$) and thermal conductivity ($k(\varphi)$) across the domain are:

$$\rho(\varphi) = \rho_l H_\varepsilon(\varphi) + \rho_g(1 - H_\varepsilon(\varphi)) \tag{8}$$

$$\mu(\varphi) = \mu_l H_\varepsilon(\varphi) + \mu_g(1 - H_\varepsilon(\varphi)) \tag{9}$$

$$c_p(\varphi) = c_{p_l} H_\varepsilon(\varphi) + c_{p_g}(1 - H_\varepsilon(\varphi)) \tag{10}$$

$$k(\varphi) = k_l H_\varepsilon(\varphi) + k_g(1 - H_\varepsilon(\varphi)) \tag{11}$$

## 2.4. The Evaporation and Condensation Model

The implementation of this evaporation and condensation model has been presented in [37]. The model is verified by comparing bubble growth rates against analytical solutions. The numerical bubble release frequency is validated by comparing commonly-used experimental correlations under pool boiling conditions, and the bubble evolution and numerical bubble growth rates are validated against experimental results under flow boiling conditions. A brief introduction is provided below to help understand the evaporation and condensation model used in the paper.

This evaporation and condensation model is designed to resolve 3D interfaces in complex geometries represented by unstructured meshes. This unique capability allows the investigation of boiling phenomena under various conditions at lower numerical costs (by utilizing local mesh refinement in the bubble growth region) compared to uniformly refined structured grids. Bubble evaporation and condensation is achieved by coupling the scalar equation which calculates the volume increase/decrease due to phase-changes with the continuity and momentum equations. In boiling simulations, the energy equation is included as part of the flow solution. The local temperature gradient is estimated to obtain the averaged heat flux through the interface according to the temperature distribution around each individual bubble (vapor temperature is set to be at saturation). The volume increase based on this average heat flux is uniformly added into the mass conservation equation as a volumetric source term controlling the growth rate. This approach ignores the effect of local temperature variation on the bubble growth. The advantage of this averaging approach is that it makes the simulations more stable and allows for smooth bubble growth and condensation even during localized temperature fluctuations.

In the estimation of the average temperature gradient for individual bubbles, the evaporation and condensation model is coupled with a Bubble Tracking Algorithm (BTA) [38]. BTA introduced a marker/ID field in PHASTA to identify and track each individual bubble in the domain as shown in **Figure 1.** The nodes within the region of interest are colored by the corresponding bubble ID while the rest of the domain is marked by zero ID value. The region of interest consists of two parts: the bubble region (to collect the bubble-related information) and a near interface liquid shell (to collect local liquid information). The temperature gradient information is collected within the "liquid shell" region. After coupling with BTA, the evaporation and condensation model can estimate the average temperature value for every bubble and achieve different growth rates for each bubble at various thermal conditions.

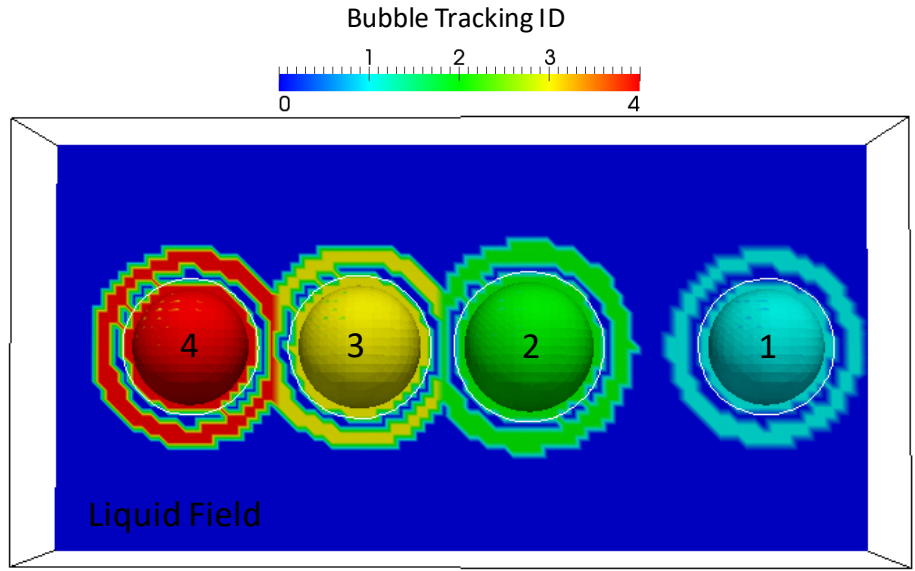

**Figure 1. Four-bubble simulation colored by bubble tracking marker field (zero value indicates liquid field).**

## 2.5. The Contact Angle Control Model

The contact angle effect quantifies the wettability of a solid surface by a liquid. It plays an essential role in determining the characteristics of boiling phenomena such as nucleation site density [6], bubble departure diameter [39], and bubble release frequency [5]. The value of the contact angle is determined by the solid surface, liquid and vapor properties at micrometer- and nanometer-scales. Interface tracking simulations cannot directly resolve such small scales. Therefore, a contact angle control model is introduced to represent the contact angle effect within millimeter-scales (the scale-length of interface tracking simulations). The implementation and verification of such a contact angle control model is developed in [36]. The coupling between contact angle control and the evaporation and condensation model is presented in [37].

This contact angle control model is implemented as a subgrid force model in PHASTA. The contact angle effects are represented by the value of a prescribed target contact angle in the algorithm. As a computational fluid dynamics model approach, PHASTA cannot directly simulate the properties of surface materials. Instead, PHASTA introduces a target contact angle measured for certain materials and conditions as a model input to represent the surface material effects on the apparent contact angle. The model implemented in the solver focuses on how to achieve this prescribed contact angle value and maintain the correct apparent contact angle in the simulations. The subgrid control force is applied when the current contact angle deviates from the desired value (or range of values) and decreases to zero when the current contact angle reaches the desired value. The advancing and receding contact angles are treated separately in consideration of the lateral movement of the bubble.

## 3. SIMULATIONS OF BOILING PHENOMENA

The simulations presented herein demonstrate the modeling capabilities of capturing bubble-bubble interactions in nucleate boiling phenomena. Various mechanisms are considered in the simulations. The phase-change mechanism is represented by the evaporation and condensation model. The contact angle as an essential characteristic of boiling departure diameter and frequency is considered by the contact angle control algorithm. Information on bubble dynamics, including bubble trajectory, growth rate, departure time, etc., is directly resolved using the DNS/ITS approach and this information is obtained for individual bubbles using the bubble tracking algorithm [38]. Two prescribed nucleation sites are placed on the bottom wall of the domain. In the study of nucleation site distance effects, a dimensionless cavity spacing $S/D_d$ is utilized to characterize the distance between neighboring sites. The bubble growth and departure within the "independent" region ($S/D_d > 3$) and the "prohibitive" region ($1 < S/D_d < 3$) are simulated. The effects of neighboring bubble size and contact angle are also investigated in the "prohibitive region". When a large bubble is forming in the nucleation site or the value of contact angle varies, it may have an influence the local hydrodynamic and thermal condition and affect bubble growth in the adjacent sites.

## 3.1. Pool Boiling with Two Nucleation Sites

The domain and mesh design of the pool simulations are shown in **Figure 2**. Two identical cavities are prescribed on the bottom wall of the domain. The distance between nucleation sites is 12 $mm$, which is designed to be the "independent" region. Two small vapor bubbles with the same initial radius ($r = 0.56$ mm) are introduced to the nucleation sites ($D = 1.0$ mm). The vapor phase is saturated (100°C, 1.0atm) under atmosphere pressure. The bottom wall is heated with a constant heat flux ($q'' = 50$ W/m$^2$) while the top wall is subcooled at 90°C. The initial temperature profile is linearly interpolated using the top wall temperature (90°C) and initial bottom wall temperature of 103°C. A unique 'vents' design is introduced in the top region of the domain to compensate volume changes due to significant density differences and to avoid the possible occurrence of backflow. A constant velocity inlet boundary condition is applied to the two vents while a natural boundary condition is applied to the top boundary of the domain as outlet. Gravity acts along the vertical direction of the domain. A small initial velocity is applied for numerical stability.

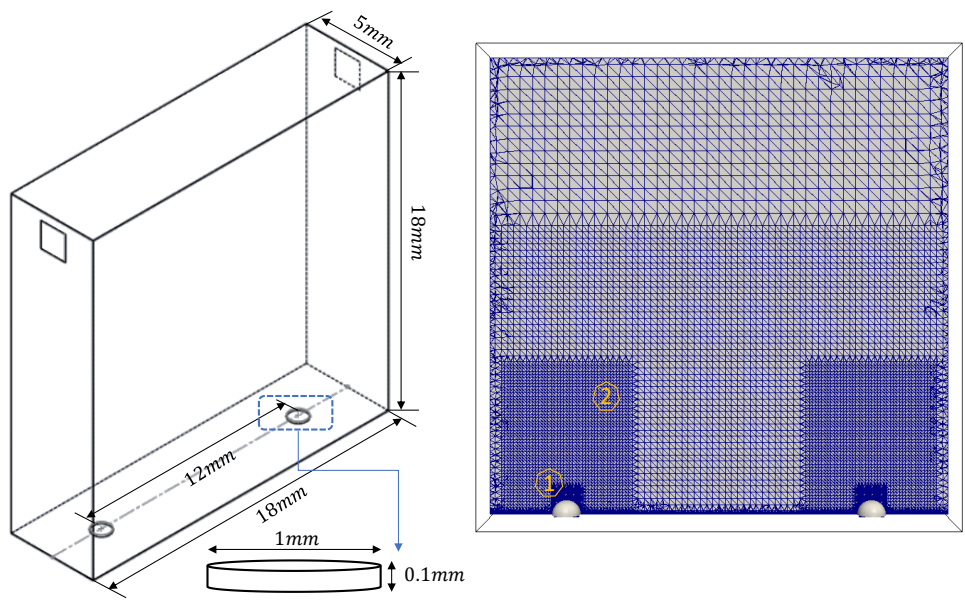

**Figure 2. The domain and mesh design of pool boiling simulation with two nucleation sites ($S = 12$ mm).**

The total resolution of the simulation is 1.8 million unstructured tetrahedral elements (0.39 million mesh nodes). The system computational time of such boiling simulations is around 10 hr utilizing 256 processing cores on a local AMD Opteron-based cluster. The mesh resolution of each refinement region is estimated based on the anticipated bubble size to ensure resolving the bubble shape and the interface accurately. The finest mesh region (Region 1) around the bubble ensures sufficient resolution to resolve the bubble shape and interface accurately. There are initially 28 elements across the bubble diameter. As the bubble grows, it is better resolved. When the bubble moves upward to the second refinement region (Region 2), its size is larger in Region 2 than Region 1 because of the evaporation. Therefore, the slightly coarser mesh is sufficient to maintain the number of elements across larger, departed bubble diameters. Moreover, the ratio of element sizes between neighboring refinement regions is no larger than two to avoid numerical instability in the simulation except for the coarsest region which is far away from the region of interest. A prism boundary layer mesh is applied in the near wall region to capture the sharp temperature gradient as well as to obtain accurate wall normal vector directions and magnitudes for the contact angle sub-grid model. The boundary layer mesh is parallel to the wall. Therefore, the vertical height of the local elements can represent the wall normal in the simulation.

The simulation results of nucleate boiling from two sites are presented in **Figure 3**. The bubble growth rates are identical and independent from each other. The observed bubble departure diameters are 3.4 mm for both bubbles. The dimensionless cavity spacing $S/D_d$ is equal to 3.5, which is within the "independent" region [8]. This simulation serves as the base case in comparison to scenarios where the nucleation site distances, or neighboring bubble sizes vary.

Two bubbles depart from the nucleation sites on the bottom wall around the same time (at 56 ms). A series of snap shots show the bubble movement and the bubble ID field (**Figure 3**). The thermal boundary layer can be seen developing around the bubble interface. Bubble growth is observed while they sit on the heated wall. As the bubble size increase, departure occurs when the buoyancy force exceeds the surface tension force and quenching effect is observed in the wake region of the bubbles. After the first two bubbles lift off from the wall, a small portion of vapor is trapped in both nucleation sites and serves as a nucleation seed for the second generation of bubbles. The departed bubbles condense quickly when they reach the subcooled region. As the first two bubbles depart from the wall, new bubbles are quickly identified by BTA. The bubble forming on the left nucleation site, for example, increases from 'ID 2' to 'ID 4' after it lifts off from the wall while new bubble at the same nucleation site is identified with 'ID 2'. The average local temperature gradient is collected for each bubble according to the bubble IDs. It is noted that the bubble growth time and waiting time is naturally determined by the force balance and heat transfer mechanisms, which are explicitly resolved in the simulations. With the boiling model presented, no additional parameters are required for the nucleating process.

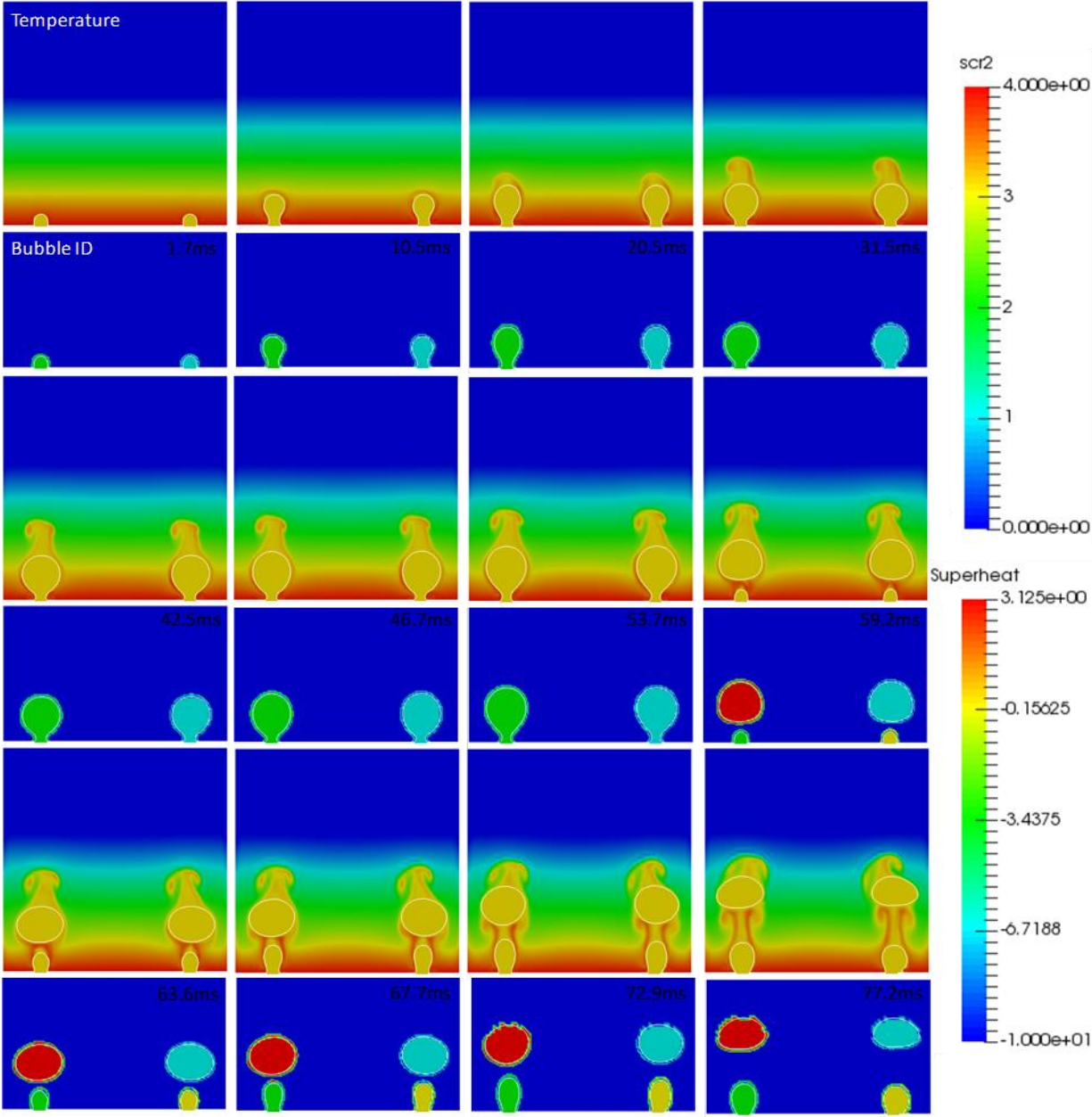

**Figure 3. Center slice of the domain with temperature distribution and bubble ID tracking.**

### 3.1.1. The Effect of Nucleation Site Distance

In the previous section, the nucleation sites are relatively far away from each other. The bubble growth and departure is independent and identical under the same hydrodynamic and thermal conditions in the domain. The effect of nucleation site distance is investigated in this section. The distance between neighboring sites is narrowed from 12 mm to 4 mm while other initial and boundary conditions remain the same as in the base case. The domain and mesh design are shown in **Figure 4**. This simulation is carried out on 1.6 million unstructured elements (0.36 million nodes), which utilized 256 processing cores.

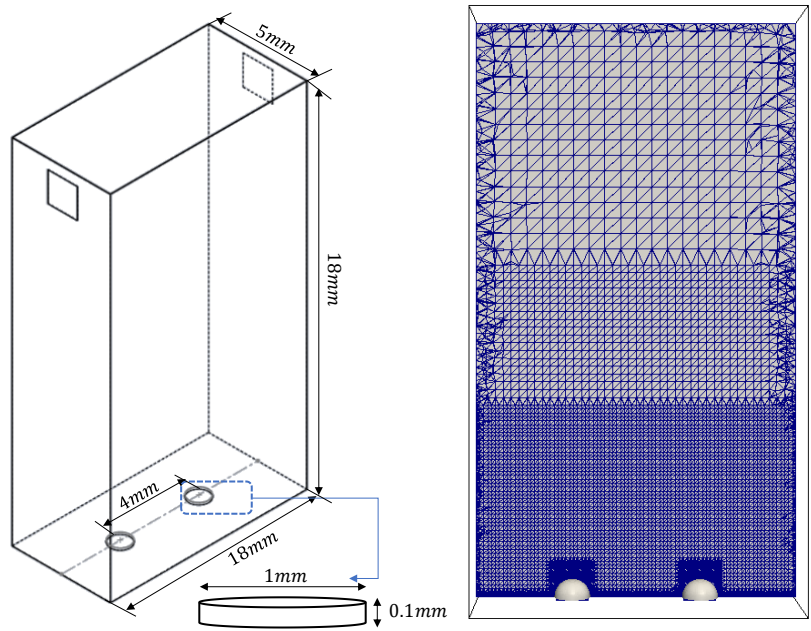

**Figure 4. The domain and mesh design of pool boiling simulation with two nucleation sites (S = 4 mm).**

In the early stage of nucleate boiling, the bubble growth rates at both nucleate sites are identical. As the bubble size increases, bubble-bubble interactions are observed in the simulation. The first two bubbles depart from the wall around the same time and quickly coalesce into one larger bubble. The lateral bubble merger near the nucleation site is observed in the simulation (**Figure 5**). Some cooler liquid is trapped between the two nucleation sites when the bubble coalescence occurs. The large coalesced bubble disturbs the local fluid conditions and interacts with new bubbles growing at the nucleation sites. As the large bubble lifts off and reaches the subcooled region, bubble condensation becomes the dominant heat transfer mechanism and shrinks the bubble quickly. The bubble departure diameter extracted from the simulation is 3.6 mm. The dimensionless cavity spacing is $S/D_d = 1.11 < 3$. According to Calka and Judd [8], the bubble formation at one nucleation site has a negative effect on the bubble formation at an adjacent nucleation site. Bubble departure is delayed compared to the base case in **Figure 6**. The simulation is plausible and consistent with the 'prohibitive' theory. It is noted that there is an small delay in **Figure 6** compared to **Figure 5**. The discrepancy comes from the data collection method -- Bubble Tracking Algorithm (BTA). As mentioned in Section 2.4, the bubble tracking algorithm collects bubble information like bubble radius for each bubble. In **Figure 6**, the radius of Bubble ID 2 is chosen to represent the bubble departure frequency. However, the bubble tracking algorithm detects new bubble at certain frequency (e.g. every 80-time steps). Before the new bubble ID is applied, bubble ID 2 and bubble ID 3 are still considered as one bubble. That leads to the discrepancy observed in **Figure 6**. As part of the numerical method, we can increase the bubble detection frequency, but it requires additional computational time. The detection frequency balances the accuracy and efficiency. Meanwhile, a more sophisticated collection algorithm will be investigated as part of the future work. In the current simulation, the single bubble growth rate plot in **Figure 6** (and later **Figure 9**) is only utilized to better present the comparison between different conditions. The accurate bubble departure time is determined with simulation screenshots.

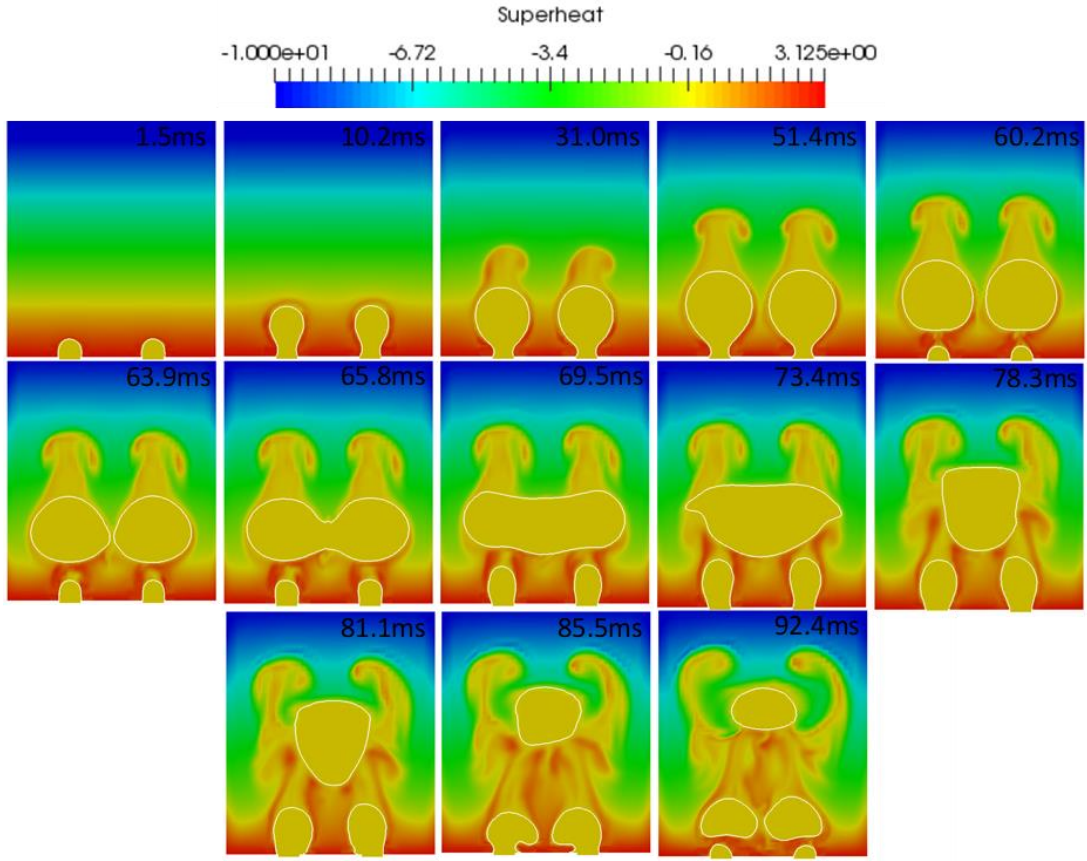

**Figure 5. Center slice of the domain with temperature distribution showing bubble evolution and lateral bubble merging.**

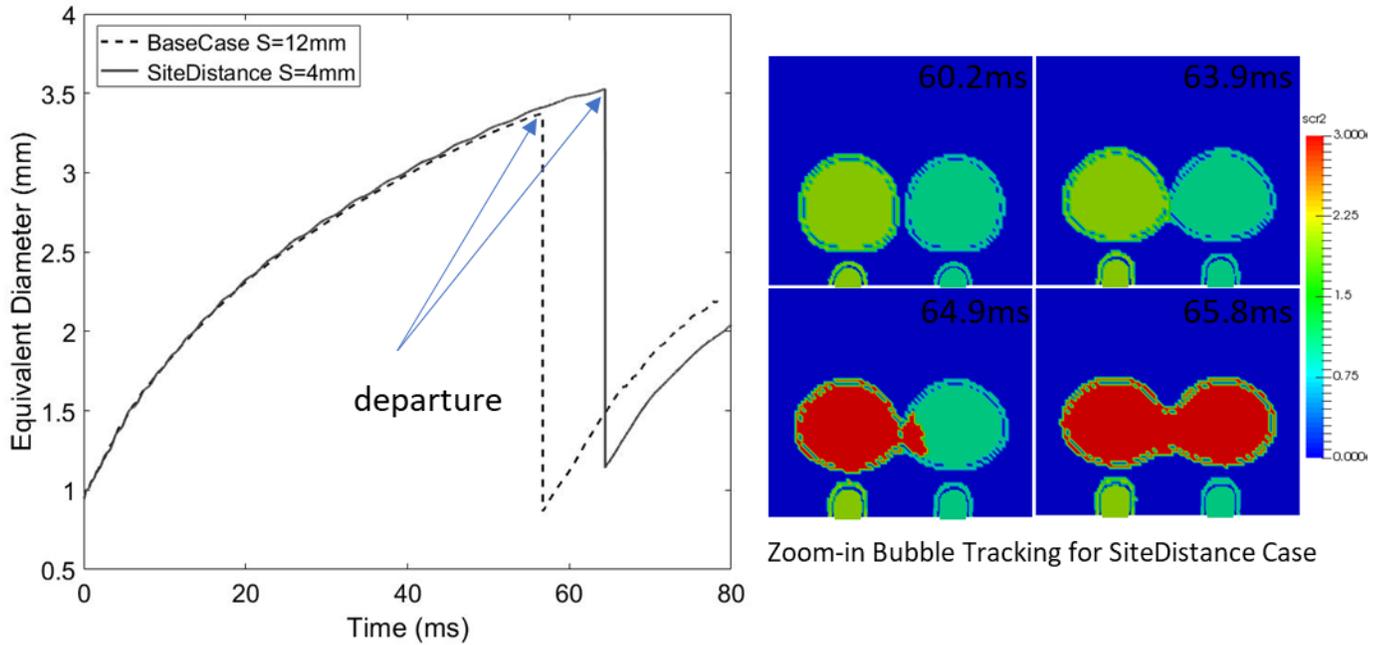

**Figure 6. Growth rate of Bubble ID 2 for site distances of $S = 12$ mm and $S = 4$ mm.**

### 3.1.2. The Effect of Neighboring Bubble Sizes

The effect of neighboring bubble sizes is investigated in this section. A large bubble is initialized in one of the nucleation sites while other initial and boundary conditions remain the same as in Section 3.1.1 (**Figure 7**). This simulation uses 1.7 million unstructured elements (0.37 million mesh nodes).

Given the same moderate heating condition, the bubble with a larger surface area requires more energy transfer through the interface, which results in less superheat in the surrounding liquid. The distance between nucleation sites is $S/D_d = 1.11 < 3$, which is smaller than the cavity spacing requirement of an "independent" region, therefore the bubble forming at the right nucleation site is affected by its neighboring bubble, as shown in **Figure 8**. The bubble growth rate and departure timing are compared with the base case simulation in **Figure 9**. It is observed that the existence of the larger initial bubble at the neighboring site delays the bubble departure time of the smaller bubble. Compared to Section 3.1.1, the bubble departure diameter is smaller, which results from the lower surrounding liquid temperature when a large bubble is forming at the adjacent site.

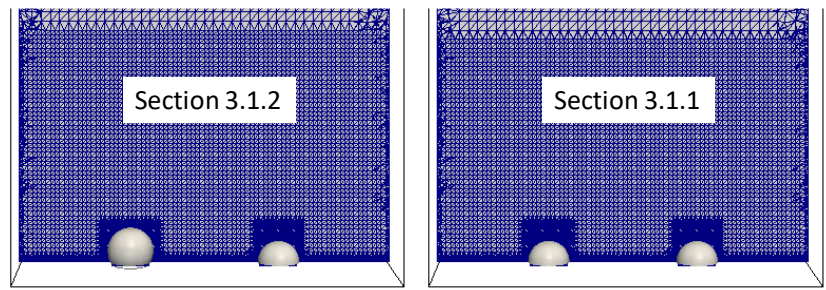

**Figure 7. The mesh design and different initial bubble sizes.**

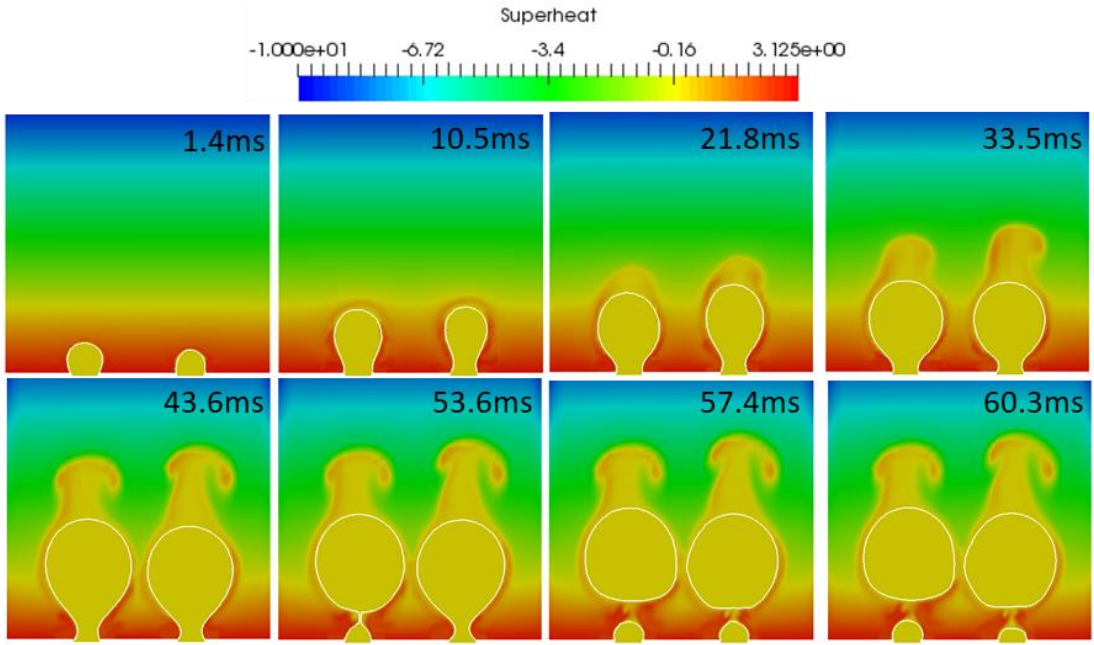

**Figure 8. Center slice of the domain with temperature distribution showing bubble evolutions with different initial radii.**

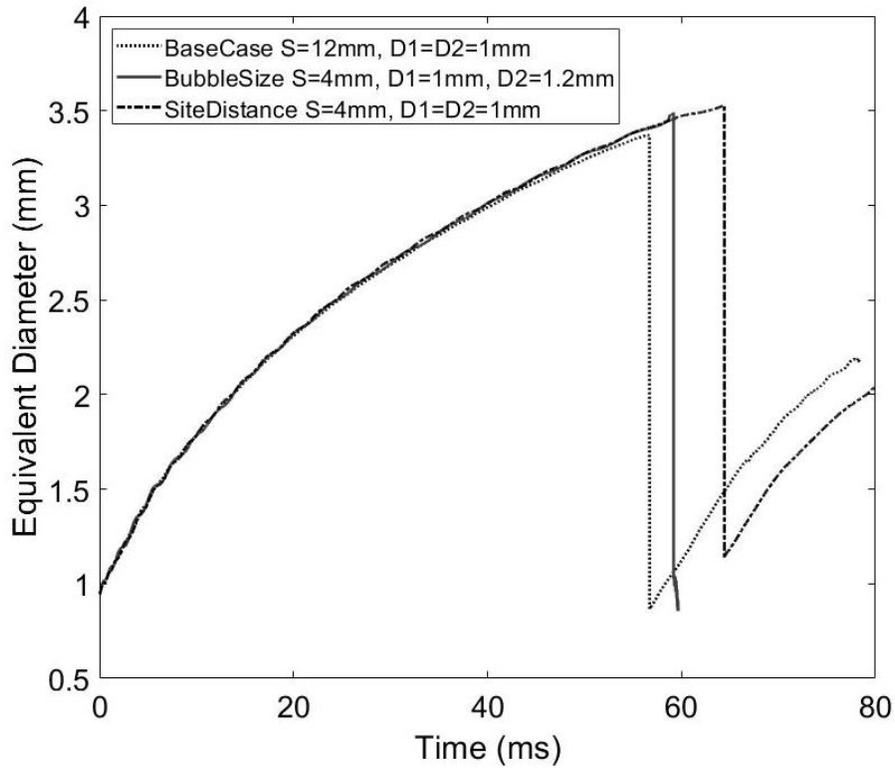

**Figure 9. Growth rate of bubble at the right nucleation site in comparison with the base case in Section 3.1.1.**

### 3.1.3. The Effect of Contact Angle

The contact angle plays an essential role in determining the characteristics of boiling. As mentioned in Section 2.5, interface tracking simulations cannot directly resolve such small scales. Instead, the material properties and surface condition are represented by the target apparent contact angle value in the contact angle control model. By adjusting the target contact angle value, we can investigate the effect of contact angles on bubble dynamics and interactions in boiling phenomena.

The domain design as well as initial and boundary conditions are the same as in Section 3.1.1 (**Figure 4**). The distance between neighboring sites (4 mm) is within the 'prohibitive' region. The effect of the target contact angle is studied by selecting the following values: 30°, 50°, and 65°, which represents 3 different surfaces on which boiling occurs. As the value of the target contact angle increases, the bubble departure time is prolonged and different bubble interactions are observed. The comparison of contact angle effect on nucleation boiling is shown in **Figure 10**.

When the target contact angle is 30°, the bubble departure occurs after 60 ms. Bubble-bubble interactions occur after they lift off the wall. As the target contact angle increases to 50°, the bubbles depart from the wall at a later time of 88 ms. Bubbles coalesce right at the nucleation site. A large bubble forms and lifts off from the wall. Two new bubbles form at the sites with trapped vapor. A target contact angle of 65° indicates relatively low wettability of the wall. The bubbles intend to attach on the wall. As the bubble diameters increase, lateral merging of bubbles is again observed. Different from the 50° case, the merged bubble forms a liquid film on the wall instead of lifting off.

The simulated bubble behaviors qualitatively agree with experimental results of nucleation pool boiling. The numerical representation of contact angle effects demonstrates a novel pathway to help design heater surface properties with desired boiling characteristics.

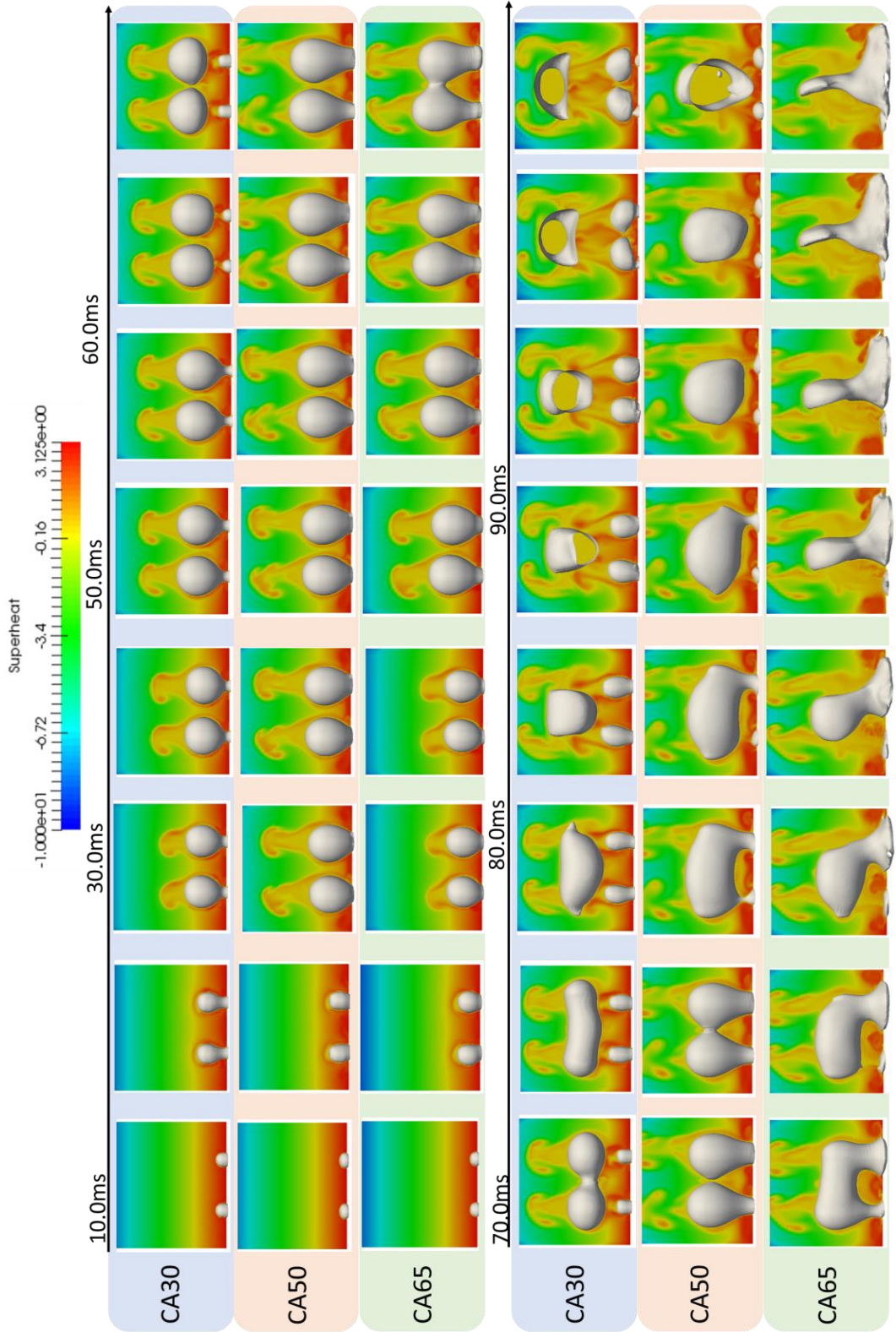

**Figure 10. Comparison of contact angle effects on nucleation boiling for three contact angle values.**

## 4. CONCLUSIONS

The interaction between bubbles forming at adjacent nucleation sites is investigated using an ITM boiling model. The influences of site distances, neighboring bubble size and contact angle effect are studied. In the 'independent' region, the bubble growth rates are identical and independent from each other while in the 'prohibitive' region, the bubble formation at one nucleation site has a negative effect on the bubble formation at an adjacent nucleation site. The neighboring bubble size has an impact on bubble formation. The contact angle representing surface condition affects bubble growth in the nearby site especially departure time.

The qualitative agreement with the nucleation site interaction model in the literature drawn from the comparison with the base case using the same initial condition while varying individual controlling variables. With this control variable method, the presented simulations only consider the first bubble cycle. Future work will extend the boiling simulation to multiple bubble cycle and a more thorough quantitative analysis of bubble statistics. The numerical bubble departure frequency will also be validated against experimental data. The simulation results presented in this paper demonstrate the potential of this ITM boiling model in helping to understand physical mechanisms of the nucleate boiling phenomenon and the flexibility to evaluate boiling behavior in complex geometries and on surfaces with different contact angle properties. This work is intended to help further bridge that gap in marrying the study of local phenomena and large-scale engineering problems.

## ACKNOWLEDGMENTS

This work is partially supported by the Department of Energy's Nuclear Energy University Program's Integrated Research Project "Development and Application of a Data-Driven Methodology for Validation of Risk-Informed Safety Margin Characterization Models", and by Consortium for Advanced Simulation of Light Water Reactors (http://www.casl.gov), an Energy Innovation Hub (http://www.energy.gov/hubs) for Modeling and Simulation of Nuclear Reactors under U.S. Department of Energy [grant number DE-AC05-00OR22725]. The last author appreciates the support of the Alexander von Humboldt (AvH) foundation Fellowship for

Experienced researchers. The solution presented here is using the Acusim linear algebra solution library provided by Altair Engineering Inc. and meshing and geometric modeling libraries by Simmetrix Inc.

# NOMENCLATURE

| | |
|---|---|
| $\vec{u}$ | Fluid velocity (m/s) |
| $H_\varepsilon$ | Smoothed Heaviside function |
| $d$ | Corrected distance field in level set method (m) |
| $t$ | Simulation time (s) |
| $\vec{f}$ | Body force density (N/m$^3$) |
| $f_s$ | Interfacial force density (N/m$^3$) |
| $T$ | Absolution temperature of the fluid (K) |
| $c_p$ | Specific heat at constant pressure (kJ/(kg · °C)) |
| $c_v$ | Specific heat at constant volume (kJ/(kg · °C)) |
| $S_{ij}$ | Strain rate tensor |
| $k$ | Thermal conductivity (W/(m · K)) |
| $\rho_l$ | Liquid density (kg/m$^3$) |
| $\rho_g$ | Gas density (kg/m$^3$) |
| $\varepsilon_l$ | Interface half-thickness in the level set equation (m) |
| $\varepsilon_d$ | Interface half-thickness in the re-distancing equation (m) |
| $\mu_l$ | Liquid dynamic viscosity (N · s/m$^2$) |
| $\mu_g$ | Gas dynamic viscosity (N · s/m$^2$) |
| $\nu$ | Liquid kinematic viscosity (m$^2$/s) |
| $\varphi$ | Level set scalar variable (m) |
| $\tau_{ij}$ | Reynolds stress tensor (N/m$^2$) |
| $\phi$ | Distance from the interface |
| $\sigma$ | Surface tension (N/m) |
| $S$ | Distance between two cavity centers (m) |
| $D_d$ | The average bubble departure diameter (m) |